\documentclass[10pt,aps,prl,floats,twocolumn,showpacs,superscriptaddress,nobibnotes]{revtex4-2}
\usepackage[svgnames]{xcolor} 
\usepackage{graphicx,epsfig}
\usepackage{times} 
\usepackage{amssymb,amsmath,multirow,rotate,color}
\usepackage{xcolor}
\usepackage[normalem]{ulem}
\usepackage{comment}

\definecolor{albicocca}{rgb}{0.98, 0.7, 0.2}
\definecolor{internationalorange}{rgb}{1.0, 0.31, 0.0}
\definecolor{giocolor}{RGB}{0, 150, 100}

\definecolor{blueresponse}{RGB}{61, 133, 198}

\usepackage{float}
\usepackage{lipsum} 
\usepackage{ragged2e}
\usepackage{soul}
\DeclareMathAlphabet\mathbfcal{OMS}{cmsy}{b}{n}

\usepackage{booktabs}

\definecolor{color1}{RGB}{109,147,140}
\definecolor{color2}{RGB}{208,98,63}
\definecolor{color3}{RGB}{222,153,97}
\definecolor{color4}{RGB}{192,60,46}
\definecolor{color5}{RGB}{31,34,57}
\definecolor{color6}{RGB}{43,86,115}

\begin{document}






\title{Genotype networks drive oscillating endemicity and epidemic trajectories in viral evolution}





\author{Santiago Lamata-Otín}
\affiliation{Department of Condensed Matter Physics, University of Zaragoza, 50009 Zaragoza, Spain}
\affiliation{GOTHAM lab, Institute of Biocomputation and Physics of
Complex Systems (BIFI), University of Zaragoza, 50018 Zaragoza, Spain}

\author{Octavian C. Rotita-Ion}
\affiliation{Department of Condensed Matter Physics, University of Zaragoza, 50009 Zaragoza, Spain}

\author{Alex Arenas}
\affiliation{Departament d’Enginyeria Informàtica i Matemàtiques, Universitat Rovira i Virgili, 43007 Tarragona, Spain}
\affiliation{Pacific Northwest National Laboratory, 902 Battelle Boulevard, Richland, Washington 99354, USA}

\author{David Soriano-Paños}
\thanks{These two authors contributed equally}
\affiliation{Departament d’Enginyeria Informàtica i Matemàtiques, Universitat Rovira i Virgili, 43007 Tarragona, Spain}
\affiliation{GOTHAM lab, Institute of Biocomputation and Physics of
Complex Systems (BIFI), University of Zaragoza, 50018 Zaragoza, Spain}

\author{Jes\'us G\'omez-Garde\~nes}
\thanks{These two authors contributed equally}
\affiliation{Department of Condensed Matter Physics, University of Zaragoza, 50009 Zaragoza, Spain}
\affiliation{GOTHAM lab, Institute of Biocomputation and Physics of
Complex Systems (BIFI), University of Zaragoza, 50018 Zaragoza, Spain}
\affiliation{Center for Computational Social Science, University of Kobe, 657-8501 Kobe, Japan}

\date{\today}


\begin{abstract}

Rapidly evolving viruses use antigenic drift as a key mechanism to evade host immunity and persist in real populations. While traditional models of antigenic drift and epidemic spread rely on low-dimensional antigenic spaces, genomic surveillance data reveal that viral evolution produces complex antigenic genotype networks with hierarchical modular structures. In this study, we present an eco-evolutionary framework in which viral evolution and population immunity dynamics are shaped by the structure of antigenic genotype networks. Using synthetic networks, we demonstrate that network topology alone can drive transitions between stable endemic states and recurrent seasonal epidemics. Furthermore, our results show how the integration of the genotype network of the H3N2 influenza in our model allows for estimating the emergence times of various haplotypes resulting from its evolution. Our findings underscore the critical role of the topology of genotype networks in shaping epidemic behavior and, besides, provide a robust framework for integrating real-world genomic data into predictive epidemic models.

\end{abstract}
\maketitle

\section{Introduction}

Epidemic trajectories of long-lasting viruses exhibit a wide range of behaviors and encompass diverse variant landscapes, as observed in SARS-CoV-2 \cite{who_covid_dashboard,chen2022global,roemer2023sars,markov2023evolution}, influenza \cite{google_flu_trends,webster1992evolution,nelson2007evolution,smith2004mapping}, dengue \cite{katzelnick2021antigenic}, or rabies \cite{streicker2010host} to name a few. This diversity arises from the intricate interplay between epidemic dynamics and viral evolution, which operate on compatible time scales \cite{moya2004population}. In particular, RNA viruses evolve rapidly within the strain space, acquiring mutations that enable them to evade detection by the host immune system \cite{katzelnick2021antigenic,mittal2022structural,guan2010emergence,guzman2000escape}. Notably, antigenic escape is the primary mechanism by which rapidly evolving viruses establish endemicity \cite{soriano2024eco}. This occurs because mutations in antibody-binding regions not only allow the virus to persist within a host evading its immune response but also enable reinfections of previously recovered individuals \cite{carabelli2023sars,smith2004mapping,tuekprakhon2022antibody,van2012evasion}.
\smallskip


The long-term behavior of epidemics not only hinges on the total number of mutations accumulated in antibody-binding regions but also on how the virus explores effectively the antigenic space. Traditionally, low-dimensional representations are assumed to study analytically the impact of the structure of the antigenic space on epidemic trajectories. For instance, one-dimensional representations \cite{rouzine2018antigenic,gog2002dynamics} have shown that the interplay between immunity acquisition and viral evolution gives rise to traveling waves in strain space. However, without considering multi-dimensional representations \cite{yan2019phylodynamic,marchi2021antigenic}, it is impossible to capture speciation events determining the emergence of distinct viral lineages within the same pathogen. The latter phenomenon exemplifies how the inherent complexity of antigenic space, shaped by the vast range of potential genomic mutations, must not be overlooked.
\smallskip

Nevertheless, among the vast number of possible genomic mutations, only a small fraction of genotypes are sampled and recorded in public databases \cite{gisaid}. This is due to two primary factors: not all genetic sequences successfully overcome intra-host selection to emerge as viable (and thus observable) variants, and not all circulating strains are sequenced and registered. The traditional approach in epidemiology for tracking genotype evolution relies on phylogenetic trees \cite{nei2000molecular,langedijk2024genomic}, which infer via probabilistic models viral relationships based on large datasets of genetic sequences. There, different genotypes are associated to different lineages, all coming from the original wild type. This methodology results in a tree-like structure where each genotype comes from a unique genotype and leads to a number of offspring genotypes. While phylogenetic methods provide valuable insights into viral evolution, they present significant limitations when modeling antigenic space. Specifically, phylogenetic reconstructions do not allow loops and thus do not account for the possibility of multiple evolutionary pathways leading to the same strain, a phenomenon known as convergent evolution \cite{wagner2014genotype,williams2022immunity}.
\newpage

As an alternative, genotype networks \cite{wagner2011genotype,wagner2014genotype,manrubia2021genotypes,dabilla2024structure,r2016parallel} offer a framework that captures these concurrent evolutionary trajectories, bridging the gap between real-world genomic diversity and its representation in antigenic models. In this approach, complex networks \cite{newman2018networks,latora2017complex} are used to describe genetic sequences as nodes, with links connecting sequences that differ by a single point mutation.  Through a detailed modeling process, Williams et al. \cite{williams2022immunity} constructed a genotype network for the highly antigenic HA protein of influenza A (H3N2). Their analysis revealed the emergence of non-trivial topological properties consistent with generative models based on linear preferential attachment, highlighting fundamental differences between real genotype networks and models considering simple antigenic space embeddings. Moreover, the emergence of hierarchical modular genotype networks following viral evolution has also been recently reported for viruses affecting other hosts, such as $Q\beta$ bacteriophages~\cite{seoane2024hierarchical}. 
\smallskip

The inherent complexity of antigenic space calls for the incorporation of genotype networks in eco-evolutionary frameworks to improve our understanding on how antigenic drift shapes epidemic trajectories. To address this question, we here present a minimal multi-strain epidemiological model that capture contagion, immune dynamics, and mutation through antigenic genotype networks.
 This model builds upon previous multi-strain frameworks \cite{castillo1989epidemiological,andreasen1997dynamics,kamo2002effect,ferguson2002influence,minayev2009improving,kucharski2016capturing,williams2021localization,gog2002dynamics} but reduces the unnecessary proliferation of epidemiological states by mapping immunity acquisition for each strain in the shortening of their associated infectious period. In this framework, the genotype network governs how the virus mutates across the genotype space and determines how hosts infected by one strain build partial immunity to other variants with closely related antigenic properties. Furthermore, our model accounts for the infection history of each host by introducing a memory term. This feature, present on the literature using low-dimensional embeddings of the antigenic space~\cite{gog2002dynamics,rouzine2018antigenic,marchi2021antigenic}, represents a significant improvement over previous Markovian models \cite{williams2021localization,williams2022immunity} considering genotype networks, as these models assume that immune response is only determined by the last variant contracted by the host.

\medskip

With our evolutionary epidemic model in place, 
we: (i) recover antigenic escape dynamics consistent with previous studies,
(ii) demonstrate that the topology of the genotype network alone can determine whether endemic trajectories consist of persistent seasonal waves or evolve into steady dynamics, (iii) uncover the complementary roles of mutant swarms and cross-immunity in sustaining infections, and
(iv) provide a fair reconstruction of epidemic trajectories on real-world genotype networks. We round off the manuscript by discussing the implications of the former results and the research avenues that can be addressed in the future.
\medskip

\section{Results}

\begin{figure*}[t!]
\centering\includegraphics[width=0.875\linewidth]{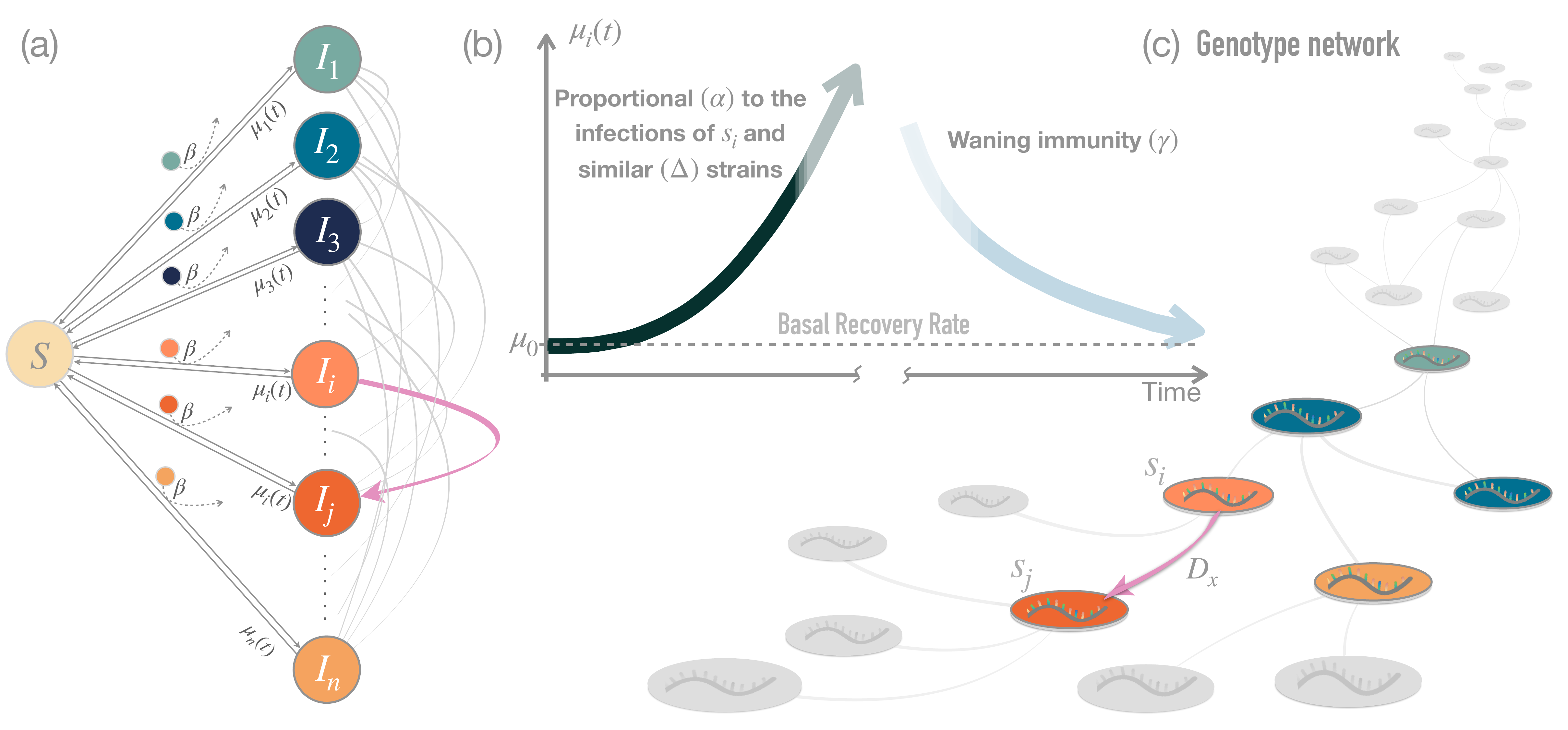}
\caption{\textbf{Schematic illustration of the SIMS dynamical model}. Panel (a) depicts the compartmental structure, where Susceptible individuals (S) can transition to one of $n$ Infectious compartments ($I_i$), each corresponding to a distinct strain $s_i$ ($i=1,\dots,n$). Infection occurs at a rate $\beta$ upon contact with an infectious individual carrying strain $s_i$. Infected individuals recover and return to the Susceptible compartment at a strain-dependent rate $\mu_i$. Panel (b) illustrates the adaptive immune response at population level: the recovery rate $\mu_i$ increases proportionally to the global immune pressure against strain $s_i$ and antigenically similar strains, with proximity in the genotype network modulated by the parameter $\Delta$. Counteracting this growth, immunity wanes over time at a rate $\gamma$, leading to a gradual reduction in $\mu_i$ toward the basal recovery rate $\mu_0$. Panel (c) represents viral evolution through mutation. Infectious individuals carrying strain $s_i$ can transition to an adjacent strain $s_j$ in the genotype network at a mutation rate $D_x$, as indicated by the colored links in panels (a) and (c). This mechanism establishes a feedback loop between contagion and mutation dynamics.}
\label{fig:1}
\end{figure*}



\subsection{Evolutionary epidemic model}\label{subsec2.1}

Modeling the interplay between contagion, immune response, and mutation dynamics presents a significant challenge, as these processes operate at distinct yet compatible spatial and temporal scales. Particularly, contagion arises from interactions between individuals in a population, immune responses occur within the host through virus-antibody interactions, and mutation is governed by the internal dynamics of viral populations.

Here, we introduce the Susceptible - Infectious - Mutation - Susceptible (SIMS) model, a compartmental framework that integrates all three processes into a unified approach. In the SIMS model, each individual belongs to either the Susceptible (S) compartment or one of $n$ distinct Infectious (I$_i$) compartments, where each I$_i$ represents an infection with strain $s_i \in \mathcal{N}$. Our model considers that intra-host viral dynamics occur on a much faster timescale than epidemiological spread. Consequently, the model neglects intra-host viral diversity and assumes that the infectious state of each individual is defined by a single strain.

\medskip

The structure of the SIMS model is illustrated in Fig. \ref{fig:1} and its main processes are described as follows:
\medskip

\noindent{\bf Contagion}: As shown in Fig.~\ref{fig:1}.a susceptible (S) individuals contract an infection upon direct contact with Infectious (I$_i$) individuals infected with strain $s_i$, which has an associated infectivity rate $\beta$. Upon infection, an S individual transitions to the I$_i$ compartment, acquiring the viral genotype $s_i$ of the infecting host. We also introduce a recovery rate $\mu_i$(t) governing how individuals infected by strain $i$ transit to the Susceptible state at time $t$.


\medskip


\noindent{\bf Immune response}: Unlike other epidemic frameworks, the recovery rate $\mu_i$ is not constant over time in the SIMS model. Instead, the recovery rates $\left\{\mu_i (t)\right\}$ store the infection history of the population. In particular, we assume $\mu_i$ to represent the immune response built in the population as a response to the proliferation of strain $i$ (see Fig.~\ref{fig:1}.b). Namely, there is a disease-free value $\mu_0$ corresponding to the baseline immune response existing in the population. Infection by strain $i$ enhances host immunity against that strain at a given rate $\alpha$. Moreover, individuals gain partial cross-immunity to genetically similar strains in the genotype network, with a characteristic cross-immunity length $\Delta$. Acquired immunity wanes over time with a characteristic decay rate $\gamma$.

\medskip

\noindent{\bf  Mutation}: Viral evolution (see Fig.~\ref{fig:1}.c) is modeled as a diffusion process through the genotype network. Specifically, an infected individual associated with strain $s_i \in \mathcal{N}$ can mutate to a neighboring strain $s_j$ at a rate $D_x$. When this mutation happens, the individual transitions from $I_i$ to $I_j$ in the compartmental dynamics (see Fig.~\ref{fig:1}.a), creating a feedback loop between mutation and contagion dynamics.
\medskip

In summary, the SIMS model consists of $n+1$ compartments and is governed by six epidemiological parameters (see Supplementary Table I), capturing the interplay between contagion, immune dynamics, and mutation in a genetically diverse viral population. The dynamics of the SIMS model can be captured by the set of coupled differential Eqs. (\ref{eq:1})-(\ref{eq:2}), described in Methods. These equations yield the temporal evolution of the fraction of population infected by each strain $i$, $\rho_i(t)$, and the global prevalence of the disease, $I(t)$ (defined in Eq. (
\ref{eq:I_tot})), hereinafter used as the principal metrics to characterize our epidemic trajectories.
\medskip

Supplementary Fig.~1.a shows that the SIMS model for a single strain represents a versatile framework that allows reproducing the  epidemic trajectories generated by multiple standard compartmental models. Namely, SIS-like dynamics are generated when neglecting the stimulation of the immune response i.e. $\alpha=0$, whereas SIR-like (SIRS-like) dynamics occur when the disease confers long-lasting (temporal) immune memory, i.e. $\gamma=0$ ($\gamma\neq0$) with $\alpha\neq 0$. 
Subsequently, in Supplementary Fig 1.b-c,  we extend the analysis to multi-strain dynamics connected through a linear chain in the antigenic space. Considering long-lasting immune memory, i.e. $\gamma=0$, the model retrieves well-known phenomenology reported by previous eco-evolutionary frameworks assuming low-dimensional antigenic spaces \cite{rouzine2018antigenic}. Namely, we observe SIRS-like trajectories generated by the antigenic drift of the virus through a travelling-wave solution across the antigenic space. To round off the description of the SIMS model, in the Supplementary Material, we derive analytically the expression for both the stationary prevalence (further analyzed in Supplementary Fig. 2) and the epidemic threshold. Likewise, in Supplementary Figs. 3 and 4 we explore regions of the parameters space yielding unexpected dynamical regimes such as genotype re-emergence and transitory chaotic-like behaviors.

\begin{figure*}[t!]
\centering\includegraphics[width=\linewidth]{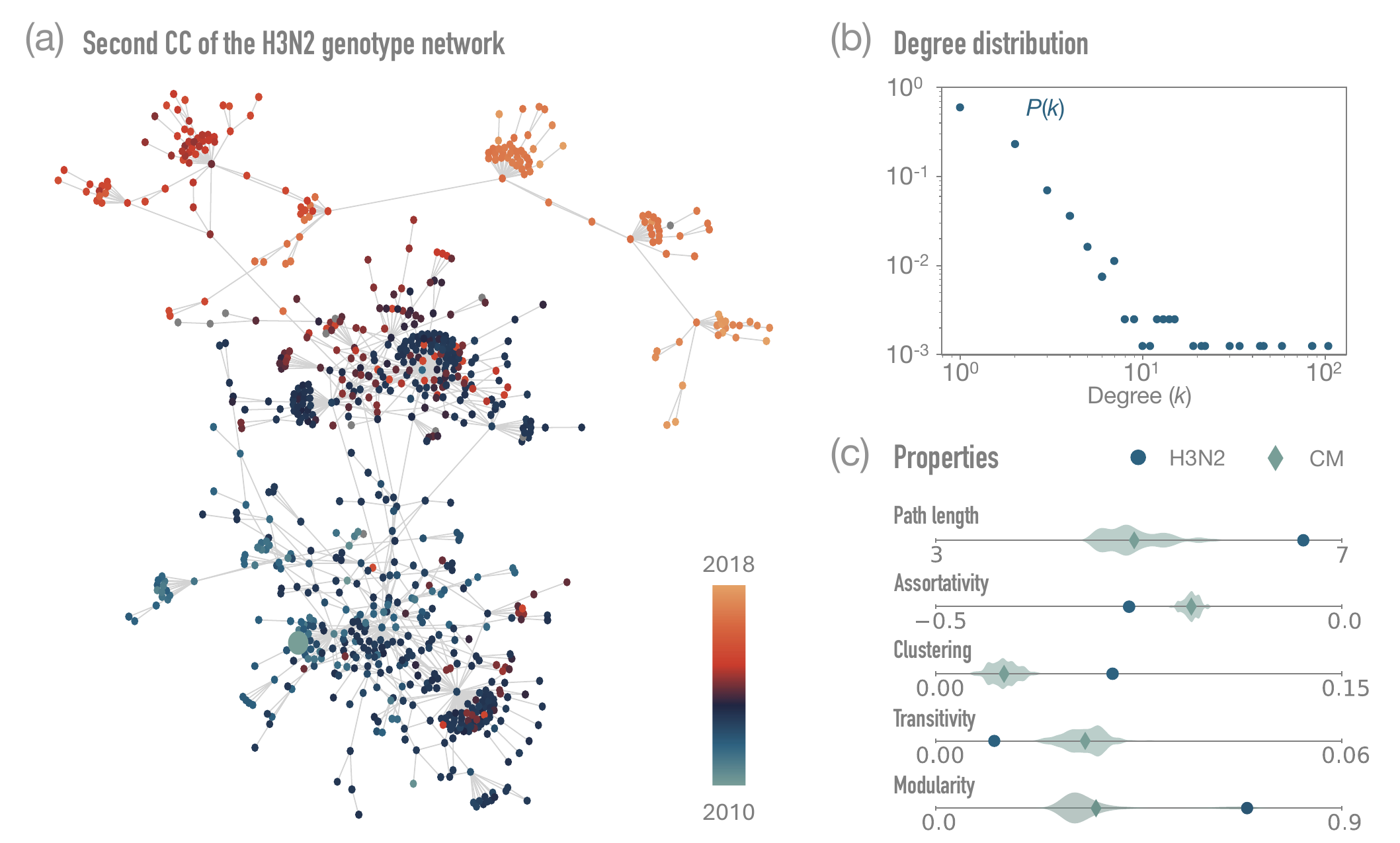}
\caption{\textbf{Structural properties of genotype networks.} (a) Second connected component of the Influenza A genotype network \cite{williams2022immunity}, where the color-code represents the sampling year. (b) Degree distribution of the second component of the Influenza A genotype network. (c) The blue dots correspond to the values of some structural properties of the second component of the Influenza A genotype network, while the violin plot corresponds to the values obtained for an ensemble of 100 networks sharing the same degree sequence. More details on the crafting can be found in the Methods. 
}
\label{fig:2}
\end{figure*}

\subsection{The structure of the antigenic space shapes epidemic trajectories}

Once presented the SIMS model, we are interested in exploring how the complex structure of genotype networks governs epidemic dynamics. To tackle this question, we should {\em i)} decipher what makes real genotype networks different from low-dimensional or randomized representations of the antigenic space and {\em ii)} understand how these features alter the behavior of epidemic trajectories.


\subsubsection{The Influenza A genotype network}

The influenza A (H3N2) genotype network~\cite{williams2022immunity} consists of multiple connected components, with eight of them containing at least 130 nodes. Fig. \ref{fig:2}.a shows the second-largest connected component, which exhibits a rich and complex topology. Fore example, its degree distribution, displayed in Fig. \ref{fig:2}.b, follows a long-tailed pattern. This feature was previously noted by Williams et al. \cite{williams2022immunity} when analyzing the largest connected component, who also identified stable clustering and negative assortativity, two structural characteristics that align with generative models based on linear preferential attachment. These observations suggest the presence of non-trivial processes governing the growth and evolution of genotype networks.
\medskip

To further investigate whether these structural features are purely a consequence of the connectivity distribution, we conduct a comparative analysis between the real genotype network and its randomized counterparts. Specifically, in Fig. \ref{fig:2}.c, we examine the most relevant structural metrics of the second-largest connected component and compare them to an ensemble of 100 degree-preserving randomized networks (see Methods for details). A similar analysis for the remaining components can be found in Supplementary Fig. 5.
\medskip

\begin{figure*}[t!]
\centering\includegraphics[width=1\linewidth]{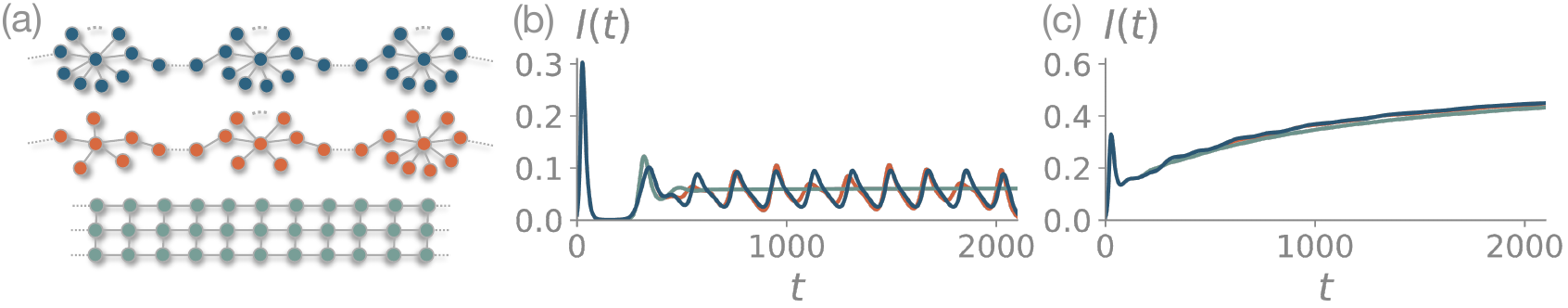}
\caption{\textbf{Steady versus seasonal endemicity.} (a) Three synthetic genotype networks: a lattice, a homogeneous concatenation of star-like swarms and a heterogeneous concatenation of star-like swarms (see Supplementary Information for details on the structures). (b) Epidemic trajectories with the memory mechanism activated ($\gamma=0$). (c) Epidemic trajectories when the time scales governing immune acquisition and loss are compatible ($\gamma=\alpha$). In all cases $\beta=0.3$, $\mu_0=0.1$, $\alpha=0.03$, $D_x=10^{-5}$ and $\Delta=3$.}
\label{fig:3}
\end{figure*}

Our results reveal substantial differences between real and randomized genotype networks. First, real genotype networks exhibit longer average path lengths than their randomized counterparts, indicating a more intricate connectivity structure. Second, higher modularity values in the real network suggest the presence of mesoscale organization, where clusters of antigenically similar genotypes, commonly referred to as {\em mutant swarms} \cite{seoane2024hierarchical}, could correspond to evolutionary lineages. Third, elevated clustering coefficients (with lower transitivity compared to randomized networks) suggest a higher likelihood of closed triangular structures, pinpointing the presence of a hierarchical organization of the networked backbone. Finally, the genotype network exhibits disassortative mixing, meaning that highly connected genotypes tend to link with those of lower connectivity. This disassortativity is consistent with the evolutionary process in which successful strains generate multiple antigenically similar offspring.

\medskip



\subsubsection{Epidemic trajectories on synthetic genotype networks}

As discussed above, the H3N2 genotype network is disassortative, presents a wide range of connectivities, and a marked modular structure. To address the impact of the complex structure of the genotype network on the epidemic trajectories, we first construct a minimal synthetic network presenting these features. In particular, we model genotype networks as a concatenation of star-like clusters (mimicking the so-called mutant swarms), where an intermediate node connects the leaves of each pair of consecutive stars (see the blue and orange structures in Fig.~\ref{fig:3} a). Regarding the epidemiological parameters, throughout the manuscript we set $\mu_0^{-1}=10$ days and $\beta=0.3$, yielding a basic reproduction number ${\cal R}_0 =3 $ (see Eq. (\ref{eq:R0}) in Methods). For the acquisition of immune response against the variants, we use an immune production rate of $\alpha=0.03$ and a cross-immunity length of $\Delta=3$. More details on the choice of these parameters can be found in the Supplementary Material. 
\smallskip

In presence of long-lasting immune memory, i.e. $\gamma=0$, the constructed synthetic genotype networks produce epidemic trajectories characterized by a series of periodic waves that lead the system to a stationary pattern of seasonal cycles. When the mutant swarms are composed of the same number of leaves (blue curve in Fig. \ref{fig:3}.b) the oscillations are regular whereas heterogeneous mutant swarms produce variability in both size and shape of the individual outbreak (orange curve in Fig. \ref{fig:3}.b). In particular, the size of a mutant swarm is proportional to the magnitude of the corresponding outbreak peak. This suggests that the emergence of mutant swarms around successful genotypes may serve to counteract cross-immunity pressure, as the simultaneous appearance of similar genotypes leads to joint outbreaks with higher overall prevalence. For further details on the trade-off between cross-immunity and star-like swarms, we refer the reader to the Supplementary Fig.~6.
\smallskip

\begin{figure*}[t!]
\centering\includegraphics[width=\linewidth]{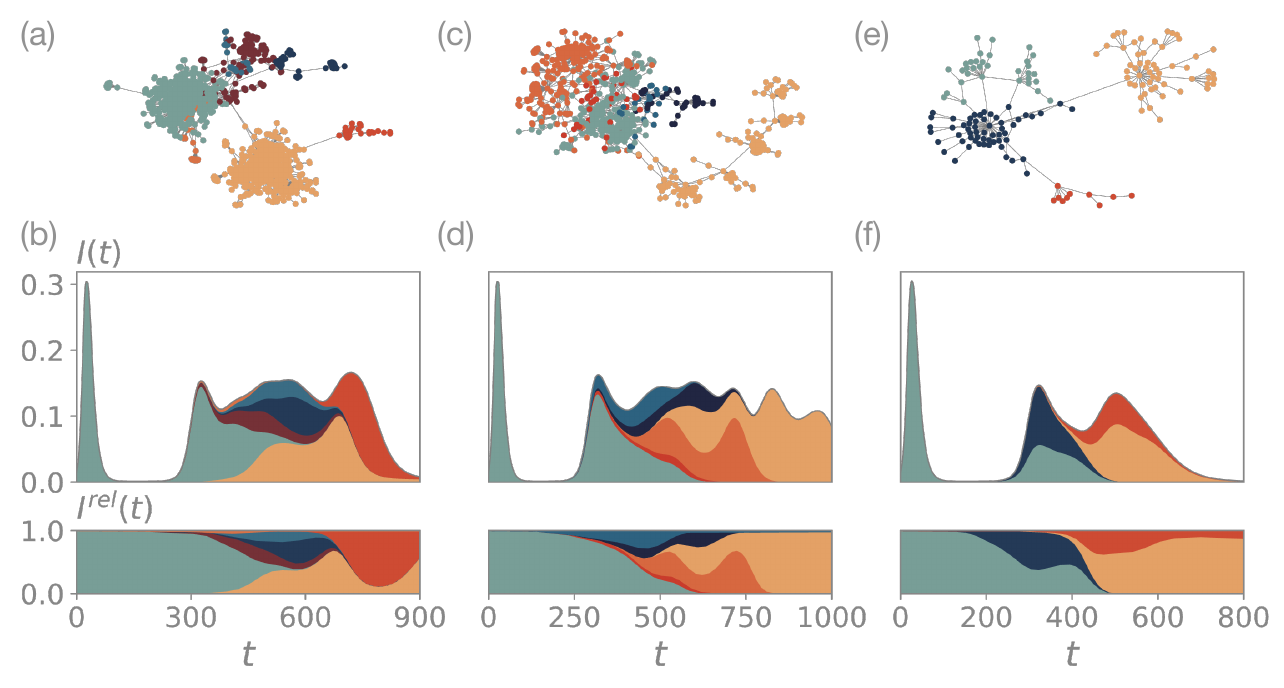}
\caption{\textbf{Epidemic trajectories in real-world genotype networks.} Graphical representations and absolute and relative epidemic trajectories for the first (panels (a)-(b)), second (panels (c)-(d)) and seventh (panels (e)-(f)) largest connected components of the INFV A (H3N2) network \cite{williams2022immunity}. The colors correspond to the mutant swarms of nodes on each of the structures. In all panels, $\beta=0.3$, $\mu_0=0.1$, $\alpha=0.03$, $\gamma=0$, $D_x=10^{-5}$ and $\Delta=3$.}
\label{fig:4}
\end{figure*}

To round off our analysis on synthetic networks, we now introduce a lattice network in the antigenic spaces, resembling the traditional low-dimensional representations of the antigenic space. In that case, we observe how the epidemic quickly converges to a steady endemic equilibrium (turquoise curve in Fig.~\ref{fig:3}.b), retrieving the well-reported results in the literature \cite{rouzine2018antigenic,chardes2023evolutionary}. Interestingly, when time scales of immune acquisition and loss are compatible $\alpha=\gamma$, Fig.~\ref{fig:3}.c shows that the epidemic trajectories for the three structures (lattice, concatenation of homogeneous swarms and concatenation of heterogeneous swarms) become macroscopically indistinguishable. Moreover, Supplementary Fig. 7 shows that SIS-like phenomenology is retrieved for the three structures if immune response dynamics are neglected. Together, our findings thus confirm that the infection history of individuals along with the complex structure of the antigenic space are responsible for the complex epidemic trajectories observed at the population level.

\begin{figure*}[t!]
\centering\includegraphics[width=0.85\linewidth]{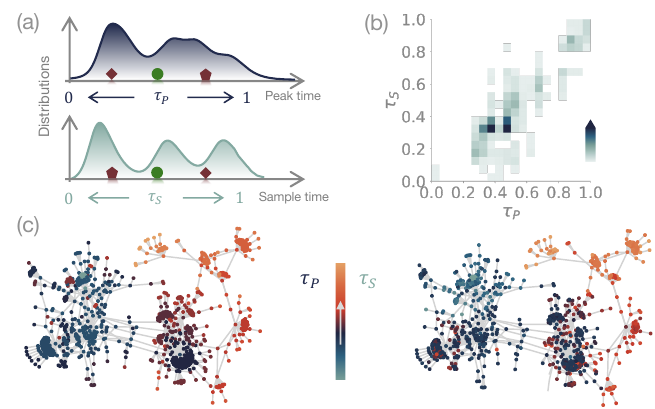}
\caption{\justifying \textbf{SIMS model reproduce partially the evolutionary trajectories of the H3N2 virus.} (a) Scheme of the methodology used to compare simulations with genotype sampling times. As an example, measurements of diamond and pentagon strains would be uncorrelated, while those of the circle strain would correlated. (b) Density scatter plot showing the correlation between peak times and sampling times for the second largest connected component. The darker the color, the more measurements correspond to that combination of values. (c) Visual comparison between the normalized sequence of peak times (left panel) and the normalized sequence of sampling times (right panel) for the second largest connected component of the INFV A (H3N2) network.  In the numerical simulations, $\beta=0.3$, $\mu_0=0.1$, $\alpha=0.03$, $\gamma=0$, $D_x=10^{-5}$ and $\Delta=3$.}
\label{fig:5}
\end{figure*}

\subsubsection{Epidemic trajectories on real world genotype networks}

After demonstrating that synthetic genotype networks can shape epidemic trajectories and the role of mutant swarms, we now turn our attention to the Influenza A genotype network constructed by Williams et al. \cite{williams2022immunity}. To follow the virus’s evolution through antigenic space, we identified the most predominant mutant swarms in each connected component of the network using community detection (see Methods). For example, Fig. \ref{fig:4}.a displays the largest connected component of the Influenza A network, with nodes in its seven communities colored differently. In Fig. \ref{fig:4}.b, the epidemic trajectory $I(t)$ is shown for this giant component, with the first sampled genotype serving as the wild type. The colored segments under the prevalence curve indicate the fraction of the population infected by each community. Initially, infections are dominated by the teal swarm, but as immunity is acquired, the antigenic escape shifts the dynamics toward the adjacent swarms. Eventually, the dynamics become dominated by the orange and red swarms, which are the most antigenically different from the wild type. For clarity, the lower panel of Fig. \ref{fig:4}.b presents the relative prevalence $I^{\text{rel}}(t)$, i.e., the fraction of infections attributable to each swarm (see Eq. (\ref{eq:I_rel}) in Methods), clearly showing the transition from the teal to the ocher communities.

In Fig.~\ref{fig:4}.c, we present the second largest component of the Influenza A network, which contains six prominent mutant swarms. Due to its more intricate topology around the wild type, the epidemic trajectory shown in Fig.~\ref{fig:4}.d is correspondingly complex, with several swarms evolving almost concurrently within the most tightly connected region of the genotype network. Nonetheless, toward the end of the trajectory, the orange swarm emerges as dominant because of its greater genetic distance from the other regions. The mutant swarm landscape captured by $I^{\text{rel}}(t)$ in this component is elaborate, as several swarms appear and fade, resembling the complex lineage patterns reported in public data repositories \cite{gisaid}.
\medskip

Similarly, Figs.~\ref{fig:4}.e–f depict the seventh largest component of the Influenza A network. This component exhibits a more hierarchical structure, similar to the synthetic concatenation of star-like clusters presented in Fig.~\ref{fig:3}. As a result, identifying mutant swarms corresponding to distinct peaks in the epidemic trajectory is more straightforward. 
\medskip


In all the cases shown in Fig.~\ref{fig:4}, epidemic trajectories are notably more complex than those derived from lower-dimensional representations, such as the linear chain shown in Supplementary Fig. 1 or the lattice used in Fig.~\ref{fig:3}. Mesoscopically, our results show how antigenic drift across genotype networks yields complex dynamic variants' landscapes whose shape and dynamics, e.g. the emergence and turnover of lineages due to antigenic fitness or the clonal interference between lineages, are reminiscent of those observed in real outbreaks. This correspondence arises from the interplay between the infection history and the trajectory of the virus across the antigenic network. Supplementary Fig. 8 shows how neglecting these features indeed leads to unrealistic variant landscapes, characterized by continuous genotype swarm alternations or the re-emergence of all lineages when cross-immunity or immune memory are not accounted for, respectively. 
\medskip

To round off our analysis, we now focus on the microscopic properties of the generated trajectories under the SIMS model. In particular, we are interested in analyzing whether the temporal dynamics of strains in our model resembles the timeline of the different H3N2 strains extracted from genomic surveillance data, i.e., the time at which the individual peak of each strain (genotype) occurred. To perform this comparison, we record the sequence of peak times of each strain $i$ in the simulation, $\tau^P_i$, and the day at which the first sample corresponding to the strain was recorded in the genomic surveillance data, $\tau^S_i$. As the global time scales of the dynamics might be different in both distributions, we normalize the values as illustrated in Figure~\ref{fig:5}.a so that $\tau^S_i=0$ ($\tau^P_i=0$) and $\tau^S_i=1$ ($\tau^P_i=1$) correspond to the first and last strain recorded in the simulations (real data) respectively. Figure~\ref{fig:5}.b reports a high correlation between the synthetic and real time series, indicating that the SIMS model can reproduce the evolutionary trajectory of the strains composing the second largest component of the antigenic genotype network of the H3N2 virus. This correspondence between the genomic data generated in silico by the SIMS model and the empirically sampled genomic data is further supported in Figs.~\ref{fig:5}.c–d, which visualize the synthetic and real epidemic trajectories over the genotype network. In both cases, nodes are colored according to the time of their first appearance in the epidemic trajectory.


\section{Discussion}

Antigenic drift has been long stated as the major driver of the sustained circulation of endemic viruses in real populations. From a dynamical point of view, the incidence of endemic communicable diseases spans a wide range of temporal patterns. Some diseases, such as tuberculosis \cite{TB} and malaria \cite{Malaria} in many regions of Africa maintain relatively stable, quasi-constant levels of incidence while others, such as influenza-like illnesses, often exhibit quasi-periodic seasonal fluctuations \cite{google_flu_trends,VRS}. While low-dimensional representations of antigenic spaces allow obtaining the former dynamical behavior, setting a formal connection between the structure of the antigenic space and the emergence of seasonal epidemics remains elusive. To solve this puzzle, here we have introduced the SIMS model, an eco-evolutionary framework where the antigenic space is explicitly represented as a complex genotype network. Another particularly salient feature of the model, compared to other frameworks relying on genotype networks, is the incorporation of a memory core that allows the system to keep track of the infection history of each host. 
\newpage 

Focusing on the H3N2 virus, we have found that real antigenic genotype networks present complex features, such as long-tailed degree distributions, high modularity, and pronounced disassortative mixing of strains which cannot be accommodated in low-dimensional antigenic spaces. We have proposed a minimal synthetic model presenting these features, assuming a complex network composed of a consecution of genotype swarms. The epidemic trajectories observed across this antigenic space exhibit oscillatory (seasonal) behavior, while assuming homogeneous lattices instead results in steady endemic epidemics. Therefore, our results show the topology of the structure of antigenic space is a critical determinant of the dynamical behavior of epidemic trajectories.
\medskip 

The SIMS model not only captures macroscopic epidemic patterns but also sheds light into the complex mesoscopic and microscopic ecological dynamics created by the antigenic drift of the H3N2 virus. When analyzing mesoscale dynamics, we have found how the interplay between immune memory, cross-immunity among strains and mutant swarms yields viral trajectories mirroring the complex lineage patterns documented in public data repositories \cite{gisaid}. From a microscopic perspective, we have found that the emergence times of the strains in the model and their first associated record in the actual genomic data are highly correlated, showing how the SIMS model can partially reconstruct the evolutionary history of the virus. 


Despite the advantages of the SIMS model, providing a full reconstruction of both epidemiological and evolutionary dynamics with our framework still remains challenging. This limitation arises because genotype networks display multiple disconnected components which typically differ by only two or three point mutations from at least one other component, indicating the possibility of unobserved genotypes linking otherwise distinct clusters~\cite{williams2022immunity}. To overcome this issue,  promising avenues for future research could involve leveraging Bayesian inference or machine learning techniques to reconstruct the complete network \cite{bonacina2024characterization,parino2025integrating} or the formulation of generative models linking the microscopic genomic structure of viruses with the topology of their antigenic spaces~\cite{koelle2006epochal}.
\medskip 

Looking forward, our work establishes a quantitative framework for integrating genomic data into epidemic models and opens promising avenues for future research. Enhancing the model by incorporating heterogeneous contact networks \cite{pastor2015epidemic,granell2024probabilistic}, host mobility~\cite{yeakel2018eco,blot2024host}, dynamic strain-specific features (such as variable infectivity \cite{soriano2024eco}), and more detailed within-host dynamics \cite{zhang2022epidemic} could improve its realism and predictive power, further bridging the gap between evolutionary and epidemiological timescales. Moreover, investigating the role of immunocompromised individuals \cite{smith2023antigenic} in facilitating extensive antigenic drift—especially for emerging pathogens like SARS-CoV-2—may yield deeper insights into viral persistence and evolution. Beyond the biological aspect of epidemics, incorporating the effect of public health policies \cite{kojaku2021effectiveness,lamata2024integrating} and the behavioral response of individuals to them~\cite{Benja2022,Saad-RoyPNAS2023,TraulsenPNAS2023} could be helpful to monitor realistic control scenarios \cite{morris2021optimal}.
\medskip

Overall, our findings underscore the critical influence of genotype network topology on epidemic trajectories and viral endemicity. This study provides valuable insights into the mechanisms by which emerging pathogens navigate antigenic space and lays a robust foundation for future research at the interface of viral genomics and infectious disease modeling. In doing so, it has the potential to inform public health interventions that account for both antigenic drift and viral circulation.

\section{Methods}
\subsection{Genotype networks}
Formally, a genotype network is defined as a graph, $\mathcal{G}=(\mathcal{N}, \mathcal{L})$, where nodes represent the $N$ distinct genetic sequences (genotypes) $s_i \in \mathcal{N}$ ($i=1,...,N$) associated to the same virus. Each sequence $s_i$ consists of a concatenation of $S$ genetic elements, i.e., $s_i={e_1^i,...,e_S^i}$. In addition, the links in the set $\mathcal{L}$ connect sequences that differ by a single mutation. The adjacency matrix $\mathbf{A}=\{a_{ij}\}$ encodes these connections, where $a_{ij}=1$ if $s_i$ and $s_j$ differ by one genetic element, and $a_{ij}=0$ otherwise. Another important matrix characterizing $\mathcal{G}$ is the out-degree normalized Laplacian matrix: 
$\textbf{L}=\{\ell_{ij}\}$, whose elements are defined as $\ell_{i j}=\delta_{ij}-a_{ij}/k_j$ where $k_i=\sum_{j=1}^N a_{ij}$ is the degree of genotype $i$. Finally, we introduce the genetic distance matrix $\mathbf{X}=\{x_{ij}\}$, whose elements are  $x_{ij}=\sum_{n=1}^{S} \delta(e_n^i, e_n^j)$, where $\delta(x,y)$ is the Kronecker delta.
Notably, if $\mathcal{G}$ is connected, $x_{ij}$ is the shortest path length between $s_i$ and $s_j$. 

Genotype networks can be constructed at different genetic scales. A node in the network may represent a genotype at three possible levels of resolution: (i) nucleotide sequences, the most fine-grained representation \cite{seoane2024hierarchical}; (ii) amino acid sequences, where nodes represent amino acids \cite{williams2022immunity}; or (iii) gene-level representations, where mutations involve entire functional units. Coarse-grained representations introduce degeneracy, as multiple nucleotide sequences can encode the same amino acid, and different genetic sequences may correspond to the same node when only a section of the genome is considered.
\medskip

Among all genomic regions, those encoding proteins involved in host-pathogen interactions are of particular interest, as they directly influence immune evasion. In particular, in this study we use the genotype network for the highly antigenic hemagglutinin protein of influenza A (H3N2), constructed in~\cite{williams2022immunity} using data from the Influenza Research Database \cite{zhang2017influenza}. This network, built at the amino acid level, connects genotypes differing by an amino acid.

\subsection{Configurational model}

To build the ensembles of 100 synthetic networks utilized in Fig. \ref{fig:2} and Supplementary Fig. 5 we have utilized the function \textit{configuration\_model} of the \textit{networkx} library in Python \cite{newman2003structure}. Through this method, for each of the connected components of the Influenza A network, we build 100 networks that have the same degree sequence (i.e. each node has the same number of connections), but with randomized connections. Note that, in case the generated structures were disconnected, we have added the least possible number of extra links to ensure connectedness.

\subsection{Structural metrics}

To analyze the structural properties of the genotype networks \(\mathcal{G} = (\mathcal{N}, \mathcal{L})\), we compute the following metrics by using the \textit{networkx} library in Python \cite{hagberg2008exploring}:


\medskip

\noindent \textbf{Average Shortest Path Length (\(L(\mathcal{G})\))} \cite{euler1741solutio,dijkstra2022note}:

\begin{eqnarray}
    L(\mathcal{G}) = \frac{1}{N(N-1)} \sum_{i \neq j} l_{ij},
    \label{eq:path_length}
\end{eqnarray}
where \(l_{ij}\) is the shortest genetic distance between strains \(s_i\) and \(s_j\), with \(L(\mathcal{G}) = \infty\) if \(\mathcal{G}\) is disconnected.


\medskip

\noindent \textbf{Degree Assortativity Coefficient $(r(\mathcal{G}))$} \cite{newman2003mixing,foster2010edge}:

\begin{eqnarray}
    r(\mathcal{G}) = \frac{\sum_{(i, j) \in \mathcal{L}} (k_i - \bar{k})(k_j - \bar{k})}{\sum_{(i, j) \in \mathcal{L}} (k_i - \bar{k})^2}.
    \label{eq:assortativity}
\end{eqnarray}
The assortativity coefficient assesses similarity in degrees (connections) between nodes, where \(\bar{k}\) is the average degree over all nodes in \(\mathcal{N}\).

\medskip

\noindent \textbf{Average Clustering Coefficient (\(C(\mathcal{G})\))} \cite{saramaki2007generalizations}:

\begin{eqnarray}
    C(\mathcal{G}) = \frac{1}{N} \sum_{i \in \mathcal{N}} c_i,
    \label{eq:clustering}
\end{eqnarray}
where \(c_i\) is the local clustering coefficient of node \(i\), defined as
\begin{eqnarray}
    c_i = \frac{2 E_i}{k_i (k_i - 1)} \quad \text{for } k_i > 1,
    \label{eq:local_clustering}
\end{eqnarray}
with \(E_i\) denoting the number of edges between the \(k_i\) neighbors of node \(i\). If \(k_i \leq 1\), \(c_i = 0\). \(C(\mathcal{G})\) quantifies the tendency of nodes to form locally dense clusters.

\medskip

\noindent \textbf{Transitivity (\(T(\mathcal{G})\))}:

\begin{eqnarray}
    T(\mathcal{G}) = \frac{3 \cdot \text{\#triangles}}{\text{\#triads}},
    \label{eq:transitivity}
\end{eqnarray}
where a triangle is a set of three nodes all mutually connected, and a triad is a set of three nodes where at least two are connected. \(T(\mathcal{G})\) evaluates the global tendency of the network to form closed triplets.

\medskip

\noindent \textbf{Modularity (\(Q(\mathcal{G})\))} \cite{newman2004finding}:

\begin{eqnarray}
    Q(\mathcal{G}) = \frac{1}{2|\mathcal{L}|} \sum_{i, j \in \mathcal{N}} \left[ A_{ij} - \frac{k_i k_j}{2|\mathcal{L}|} \right] \delta(g_i, g_j),
    \label{eq:modularity}
\end{eqnarray}
where the cardinality of the set of links \(|\mathcal{L}|\) is the number of connections, and \(\delta(g_i, g_j)\) equals 1 if nodes \(i\) and \(j\) are in the same community and 0 otherwise. Therefore, \(Q(\mathcal{G})\) measures the density of intra-community edges relative to inter-community edges.
\medskip

Note that the communities have been inferred with the Infomap method, described in the subsection below.

\subsection{Community detection algorithm}


Community detection was performed using the Infomap method \cite{rosvall2008maps,mapequation2025software} as implemented in the \textit{infomap} library in Python. This algorithm encodes the network structure as a compressed map of information flow by minimizing the description length of random walks. It iteratively partitions nodes into communities based on flow patterns, recursively refining the hierarchy until further compression is not possible.

\subsection{Effective dynamical equations of the SIMS model}

The dynamics of the SIMS model can be captured by a set of coupled differential equations. 
For simplicity, we assume a well-mixed population where all individuals are equivalent from a microscopic perspective. This leads to a mean-field description in which the relevant variables are the the probability of being susceptible, $\rho^S(t)$ and the probabilities of being infectious with each of the $n$ viral strains, $\rho^I_i(t)$ ($i=1,\ldots,n$). Equipped with this variables we are ready to formulate the set of differential equations describing
the dynamical evolution of the SIMS framework:
\begin{eqnarray}    
    \dot \rho_i^I(t) &= &\beta\rho_i^I(t)\rho^S(t)  - \mu_i(t) \rho_i^I(t) - D_x \sum_{j = 1}^n \ell_{i j} \rho_j^I(t), \label{eq:1} \\ 
    \dot \mu_i(t) &= &\alpha \sum_{j=1}^n \left(\rho_j^I(t)e^{-x_{ij}/\Delta}\right)  - \gamma(\mu_i(t)-\mu_0), \label{eq:2}
\end{eqnarray}
where the relevant parameters and matrices introduced before are present. Since the sum of probabilities associated with the state of the population at some time $t$ must fulfill:
\begin{equation}
    \rho^S(t)=1-\sum_{i=1}^n \rho_i(t)\;,
\end{equation} 
the evolution of the dynamical system is fully described by the set of $2n$ Eqs. (\ref{eq:1})-(\ref{eq:2}). 

Let us describe the terms of Eqs. (\ref{eq:1})-(\ref{eq:2}) associated with the three key mechanisms of the SIMS model. The first two terms in Eq. (\ref{eq:1}) describe the contagion dynamics following a mean-field SIS process for each viral genotype. Thus, infection occurs at rate $\beta$, while recovery takes place at a strain-dependent rate $\mu_i(t)$. Unlike the standard SIS model, where the recovery rate is constant, here it evolves dynamically according to Eq. (\ref{eq:2}) to incorporate the effects of immune response.
\medskip


In particular, Eq. (\ref{eq:2}) captures how recovery rate increases as immunity is acquired, with the gain being proportional to the prevalence of the strain at the population level $\rho_i(t)$, modulated by the immunity acquisition rate $\alpha$. This term serves as a memory core, allowing the system to keep track of past infections.
\medskip

In addition to immunity gained from direct infection, Eq. (\ref{eq:2}) also account for how individuals acquire cross-immunity from related strains. The strength of this effect depends on the genetic distance between the strains, encoded in the matrix $\mathbf{X}=\{x_{ij}\}$. Strains that are antigenically similar contribute more to immunity gain than those that are more genetically distinct. The extent of cross-immunity is controlled by the characteristic immunity length $\Delta$. When $\Delta \to 0$, immunity is strain-specific, while for $\Delta \to \infty$, all strains confer maximal cross-immunity. Finally, the last term in Eq.~(\ref{eq:2}) incorporates the effect of waning immunity over time, modeled as a decay toward the basal recovery rate $\mu_0$ with rate $\gamma$. 
\medskip

To round off,  we turn our attention again to Eq. (\ref{eq:1}). There, the last term accounts for mutation dynamics, that allows infected individuals to change infectious compartments during the course of their infection. This process is represented through the diffusion term (and thus is governed by the normalized Laplacian ${\bf L}=\{\ell_{ij}\}$) across the genotype network. The importance of this term is weighted by the mutation rate $D_x$.
\medskip

\subsection{Integration of the dynamical equations}

Epidemic trajectories are obtained by integrating Eqs. (\ref{eq:1})-(\ref{eq:2}) through a  4th order Runge-Kutta with a time step $\delta t=0.01$. There, in addition to tracking the fraction of the population infected by each viral strain, ${\rho_i^{I}(t)}$, we also use the global prevalence of the disease,
\begin{equation}
I(t)=\sum_{i=1}^{n}\rho_i^{I}(t),\;
\label{eq:I_tot}
\end{equation}
and the relative prevalence of each strain,
\begin{equation}
I^{rel}_i(t)=\frac{\rho_i^{I}(t)}{I(t)},\;
\label{eq:I_rel}
\end{equation}
as the primary metrics to characterize the disease.
\medskip
\subsection{Control parameters}

Below, we define the three control parameters utilized in the article and its Supplementary Information.
\medskip

\noindent\textbf{Basic reproduction number of SIMS model}:\\

Provided $\beta$ is the effective infectivity rate and $\mu_0$ the basal recovery rate, the basic reproduction number $R_0$ is defined as follows:
\begin{eqnarray}
    R_0=\frac{\beta}{\mu_0}.
    \label{eq:R0}
\end{eqnarray}
\medskip

\noindent\textbf{Maximum peak infectivity in SIR model}:\\

For a pathogen characterized by $R_0$, the incidence in the peak of the epidemic reads as follows \cite{lamata2023collapse}:
\begin{eqnarray}
    I_{\text{max}}=1-\frac{1}{R_0}\left[1+\text{ln}(R_0)\right].
    \label{eq:i_max}
\end{eqnarray}
\medskip

\noindent\textbf{Effective reproduction number of SIMS model}:\\

For a genotype $s_i$, provided $\beta$ is the effective infectivity rate and $\mu_i(t)$ the recovery rate, the effective reproduction number $R_i^{eff}(t)$ is defined as follows:
\begin{eqnarray}
    R_i^{\text{eff}}(t)=\frac{\beta}{\mu_i(t)}\rho^S(t).
    \label{eq:R0eff}
\end{eqnarray}

\medskip

%

\subsection{Determination of mutation rates}

Throughout this manuscript we set the mutation rate for each virus according to the empirical measurements of the literature. 
For Influenza A, the mutation rate per nucleotide per cycle takes values in the range $[10^{-4},10^{-6}]$, with a cycle lasting approximately 7 hours \cite{nobusawa2006comparison}. In particular, we have collected data from the following references:

\begin{widetext}
\begin{center}
\begin{table}[H]
\centering
\begin{tabular}{c@{\hskip 12pt}c@{\hskip 12pt}c@{\hskip 12pt}c}
\toprule
\textbf{Mutation rate per cycle} & \textbf{Mutation rate per day} & \textbf{Specifically N3H2} & \textbf{Reference} \\
\midrule
$1.5\times 10^{-5}$ & $5.14\times 10^{-5}$ & No  & \cite{parvin1986measurement} \\
$1.2\times 10^{-6}$ & $4.11\times 10^{-6}$ & Yes & \cite{nobusawa2006comparison} \\
$2.3\times 10^{-5}$ & $7.89\times 10^{-5}$ & No  & \cite{sanjuan2010viral} \\
$2.5\times 10^{-4}$ & $8.57\times 10^{-4}$ & Yes & \cite{pauly2017novel} \\
\bottomrule
\end{tabular}
\caption{Empirical measurements of mutation rates of the influenza A virus.}\label{tab:parameters}
\end{table}
\end{center}
\end{widetext}

The particular genotype network of influenza A used in the manuscript corresponds to the H3N2 type of influenza A, and only Refs. \cite{nobusawa2006comparison,pauly2017novel} account for that specific strain. Therefore, we assume as a compromise that $D_x=10^{-5}$.

\bigskip

\paragraph*{Data availability.} The Influenza A genotype network can be obtained in Ref. \cite{williams2022immunity}.
\paragraph*{Code availability.} The code is available at https://github.com/santiagolaot/SIMS-model.

\paragraph*{Acknowledgements.} S.L.O and J.G.G. acknowledge financial support from the Departamento de Industria e Innovaweci\'on del Gobierno de Arag\'on y Fondo Social Europeo (FENOL group grant E36-23R) and from Ministerio de Ciencia e Innovaci\'on (grant PID2020-113582GB-I00). S.L.O. acknowledges financial support from Gobierno de Aragón through a doctoral fellowship. A.A. acknowledges  Spanish Ministerio de Ciencia e Innovaci\'on (PID2021-128005NB-C21), Generalitat de Catalunya (2021SGR-00633), Universitat Rovira i Virgili (2023PFR-URV-00633), the European Union’s Horizon Europe Programme under the CREXDATA project (grant agreement no.\ 101092749), ICREA Academia, the James S.\ McDonnell Foundation (Grant N.\ 220020325), and the Joint Appointment Program at Pacific Northwest National Laboratory (PNNL). PNNL is a multi-program national laboratory operated for the U.S.\ Department of Energy (DOE) by Battelle Memorial Institute under Contract No.\ DE-AC05-76RL01830. D.S.P. acknowledges financial support through grants JDC2022-048339-I and PID2021-128005NB- C21 funded by MCIN/AEI/10.13039/501100011033 and the European Union “NextGenerationEU”/PRTR”.

\bibliography{biblio}

\renewcommand{\figurename}{Supplementary Fig.}
\renewcommand{\tablename}{Supplementary Table}
\renewcommand{\theequation}{S.\arabic{equation}}

\setcounter{equation}{0}
\setcounter{figure}{0}

\onecolumngrid
\newpage
\section{Supplementary Information of {\em Genotype networks drive oscillating endemicity and epidemic trajectories in viral evolution}}
\medskip

\section{Dictionary of the parameters of the SIMS model}

\begin{center}
\begin{table}[H]
\centering
\begin{tabular}{c@{\hskip 12pt}c}
\toprule
\textbf{Parameter} & \textbf{Definition} \\
\midrule
$\beta$ & Infectivity rate  \\
$\mu_0$ & Basal recovery rate  \\
$D_x$ & Mutation rate  \\
$\alpha$ & Immunity acquisition rate  \\
$\gamma$ & Waning immunity rate  \\
$\Delta$ & Characteristic cross-immunity length  \\
\bottomrule
\end{tabular}
\caption{Parameters of the SIMS model.}\label{tab:parameters_model}
\end{table}
\end{center}

\newpage

\section{Characterization of the SIMS model}

\subsection{Temporal dynamics of the SIMS model}

We first describe how the model captures basic epidemiological dynamics for a single genotype ($n=1$), where the antigenic space consists of only one strain. As shown in Supplementary Fig. \ref{fig:S2}.a for a single genotype with basic reproduction number $R_0=3$ (corresponding to $\beta=0.3$ and $\mu=0.1$), the SIMS model can exhibit three qualitatively distinct epidemic trajectories depending on immune response dynamics.
In the absence of immune dynamics ($\alpha=0$, $\gamma=0$), the model reproduces the standard SIS behavior (orange line), characterized by an initial exponential growth that stabilizes into an endemic state. When immunity acquisition is introduced ($\alpha\neq0$) while waning is negligible ($\gamma=0$), the system no longer maintains an endemic state; instead, the epidemic trajectory displays a transient wave that culminates in the absorbent state $I^{\star}=0$, mirroring the dynamics of a SIR model. In this scenario (teal line), setting $\alpha=0.03$ yields the same epidemic peak $I_{\text{max}}$ (see Eq. (13) in Methods) as the one obtained with the SIR model assuming the same basic reproduction number $R_0$. Finally, when both immunity acquisition and waning are included ($\alpha\neq0$, $\gamma\neq0$), the system converges to a nonzero endemic state following the initial outbreak (blue line), resembling the damped oscillatory behavior of the epidemic trajectories observed in SIRS models. 

\medskip

We now turn into multi-strain dynamics, where mutation interacts with contagion and recovery processes. Hereafter, we assume that immunity waning occurs on a much slower timescale than immunity acquisition ($\gamma\rightarrow 0$). Relaxing this assumption leads to genotype re-emergence and transitory chaotic-like epidemic trajectories, explored in Supplementary Fig. \ref{fig:S4} and Supplementary Fig. \ref{fig:S5} respectively. For all multi-strain simulations, the population is initially fully susceptible except for a small fraction of individuals infected with the wild-type strain ($i=1$), set as $\rho_1^I(0)=0.01$, while $\rho_i^I(0)=0$ for all $i=2,...,n$.  We set $\beta=0.3$ and $\mu_0=0.1$ so that $R_0=3$ and the immunity acquisition $\alpha=0.03$.

In the case of long-lasting immunity, once a host becomes immune against a viral genotype, the virus can only persist by mutating through the antigenic space. In Supplementary Fig. \ref{fig:S2}.b, we illustrate epidemic trajectories for different genotypes in the simplest representation of antigenic space: a linear chain network. The mutation rate is set to $D_x=10^{-5}$, inspired by empirical estimates for Influenza and other RNA viruses (see Table \ref{tab:parameters} in Methods), while cross-immunity is initially disregarded ($\Delta=0$). Unlike in the single-strain SIR-like trajectory (see Supplementary Fig. \ref{fig:S2}.a), here the virus survives beyond the first outbreak by mutating into adjacent strains. As shown in the inset of Supplementary Fig. \ref{fig:S2}.b, immunity acquired against one genotype causes its effective reproduction number $R_i^{eff}(t)$ (see Eq. (14) in Methods) to fall below 1, yet the virus persists through antigenic escape, leading to successive outbreaks of different (adjacent) genotypes. At a macroscopic level, this corresponds to a traveling wave in antigenic space, consistent with the findings of Rouzine et al. \cite{rouzine2018antigenic}, who showed that one-dimensional antigenic spaces in epidemiological models generate steady traveling waves.

\begin{figure}[h]
\includegraphics[width=\linewidth]{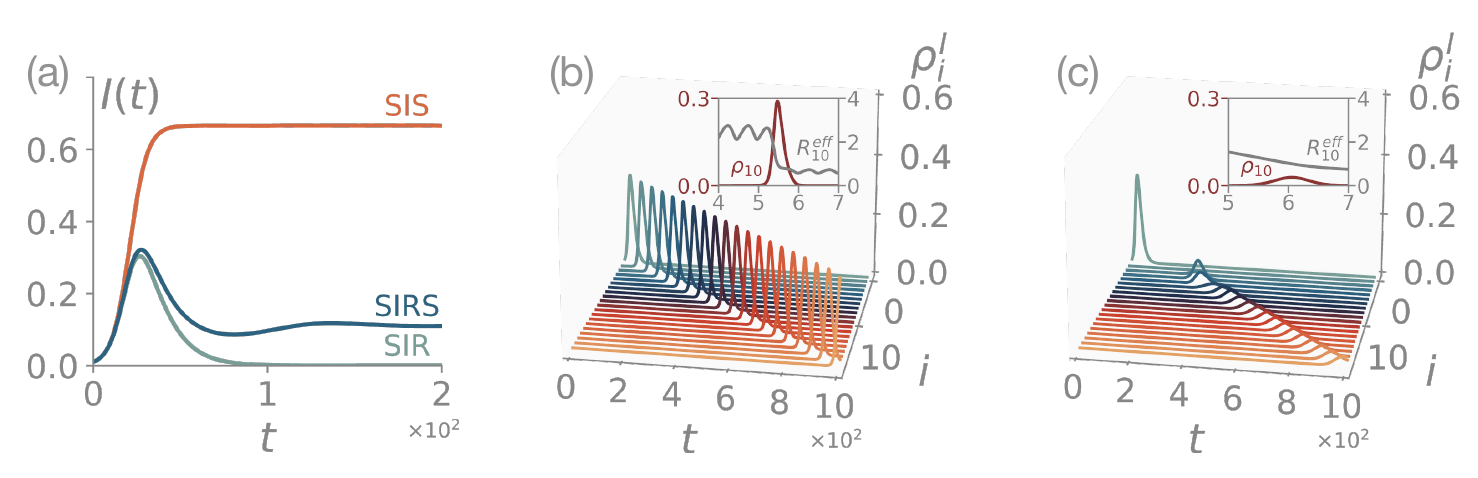}
\caption{\justifying\textbf{Dynamics of the SIMS model.} (a) The three dynamical regimes captured by the SIMS model in absence of mutations ($D_x=0)$: the endemic SIS-like regime corresponds to $\alpha=0$ and $\gamma=0$; the epidemic SIR-like wave corresponds to $\alpha=0.03$ and $\gamma=0$: and the SIRS-like
 scenario where endemicity is achieved after a first wave corresponds to $\alpha=0.03$ and $\gamma=0.02$. In all curves $\beta=0.3$ and $\mu_0=0.1$. (b)-(c) Antigenic scape by considering a linear chain as genotype network. In (b) we do not consider cross-immunity ($\Delta=0$), and in (c) we do consider $\Delta=3$. In (b)-(c) panels, the remaining parameters take the following values: $\beta=0.3$, $\mu_0=0.1$, $\alpha=0.03$, $\gamma=0$ and $D_x=10^{-5}$.}
\label{fig:S2}
\end{figure}

Cross-immunity between antigenically similar genotypes significantly influences epidemic dynamics \cite{reich2013interactions,tamura2005mechanisms}. When introducing a nonzero characteristic cross-immunity length ($\Delta=3$), we find that only sufficiently distinct mutant strains can sustain outbreaks (see Supplementary Fig. \ref{fig:S2}.c), as cross-immunity effectively suppresses the effective reproduction number so that they cannot generate an outbreak. In the linear chain representation, the genotype responsible for the second outbreak is the closest strain $s_i$ for which  $R_i^{eff}(t)>1$. Nevertheless, since $\mu_i>\mu_0$, the second outbreak has a lower impact than the initial one, failing to generate strong immunity against subsequent genotypes. This allows for the formation of a traveling wave with rescaled prevalence, resembling the SIRS-like trajectory in Supplementary Fig. \ref{fig:S2}.a.

\subsection{Epidemic diagram}

Under a mean field assumption ($\rho_i(t)=\rho(t)$ and $\mu_i(t)=\mu(t)$) and considering that $x_{ij}=0$ if $i=j$ and $x_{ij}=\langle x\rangle$ otherwise, we can rewrite Eqs. (9)-(10) as
\begin{eqnarray}    
    \dot \rho^I(t) &= &\beta\rho(t)\left(1-n\rho^I(t)\right)  - \mu(t) \rho^I(t), \label{eq:SS1} \\ 
    \dot \mu(t) &= &\alpha \left[1+ (N-1)e^{-\langle x\rangle/\Delta}\right]\rho(t)  - \gamma(\mu(t)-\mu_0).
    \label{eq:SS2}
\end{eqnarray}
In the stationary state ($\dot\mu(t)=\dot \rho(t)=0$), Eq. (\ref{eq:SS1}) presents an absorbent solution: $\rho^{\star}_0=0$. However, assuming that $\rho^{\star}\neq0$, the implicit relation $\mu^{\star}=\mu^{\star}(\rho^{\star})$ can be derived from Eq. (\ref{eq:SS1}):
\begin{eqnarray}
    \mu^\star &= &\beta (1-n\rho^{\star}). 
    \label{eq:SS3}
\end{eqnarray}
Introducing Eq. (\ref{eq:SS3}) in Eq. (\ref{eq:SS2}) and recalling that $I^*=\sum_{i=1}^N\rho_i^*=n\rho^*$ we obtain that the prevalence of the endemic phase reads:
\begin{eqnarray}
    I^{\star}=\frac{\beta-\mu_0}{\beta+\frac{\alpha}{n\gamma}\left[1+ (n-1)e^{-\langle x\rangle/\Delta}\right]}.
    \label{eq:stationary_solution}
\end{eqnarray}

In Supplementary Fig.~\ref{fig:S3}.a, we illustrate the stationary prevalence, $I^{\star}$, as a function of the basic reproduction number, $R_0$. Each curve in the epidemic diagram corresponds to a different value of the power of immune response $\bar\eta = \frac{\alpha}{n\gamma}\left[1+ (n-1)e^{-\langle x\rangle/\Delta}\right]$. Note that $\bar\eta$ combines both the rates of acquistion and loss of immune response and the cross-immunity lengths dictated by the genotype networks. As expected, when the immune system is insensitive ($\bar\eta\rightarrow0$), the model reproduces the standard SIS dynamics, with a continuous transition at the epidemic threshold ($R_0^c=1$) from an absorbent state ($R_0<1$) to an endemic state with prevalence $I^{\star}$ ($R_0>1$). Interestingly, when immune response is high ($\bar\eta\rightarrow\infty$), epidemic trajectories settle near $I^{\star}\rightarrow 0$. Consequently, without antigenic evolution, viral persistence in the population becomes unfeasible. Finally, for intermediate values of the power of immune response, the resulting prevalence curves lie between the two extreme cases. Note also that, from Eq. (\ref{eq:stationary_solution}) we realize that the endemic prevalence depends on the relevance of the power of the immune response in comparison to the basal recovery rate $\mu_0$.

Moreover, the power of immune response depends on the number of genotypes $n$. We illustrate this dependency by representing in Supplementary Fig.~\ref{fig:S3}.b the stationary prevalence, $I^{\star}$, as a function of the number of genotypes, $n$. Now, each curve in the epidemic diagram corresponds to a different value of the single-genotype power of immune response ($\eta=\alpha/\gamma$), provided $\beta=0.3$, $\mu=0.1$ and $\langle x\rangle/\Delta=100$. Notably, the stationary prevalence increases proportionally to $n$. If $n\rightarrow0$ the prevalence of the virus is naturally null ($I^{\star}=0$) and, in contrast, in the limit $n\rightarrow\infty$ the global dynamics follow the standard SIS dynamics, with $I^{\star}=1-R_0^{-1}$. This result highlights the role of viral diversity in sustaining and empowering epidemic outbreaks. 

\begin{figure}[t!]
\includegraphics[width=1\linewidth]{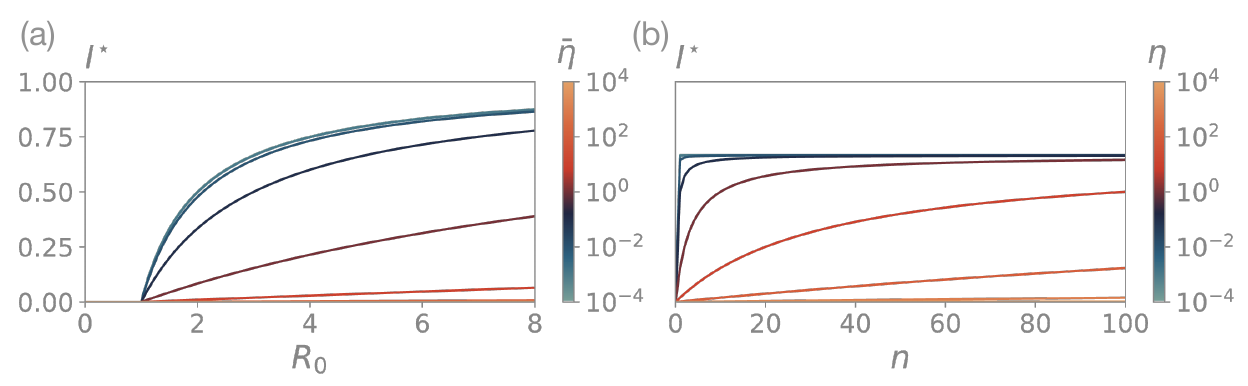}
\caption{\justifying\textbf{Epidemic diagram of the SIMS model.} (a) Global prevalence of the disease $I^*$ as a function of the basic reproduction number $R_0$ and different values of the power of immune response $\bar\eta$ (color code). In all curves we have used $\mu_0=0.1$. (b) Global prevalence of the disease $I^*$ as a function of the number of strains $n$ and the power of single-genotype immune response $\eta$. In all curves we have used $\beta=0.3$, $\mu=0.1$ and $\langle x\rangle/\Delta=100$.
}
\label{fig:S3}
\end{figure}

\subsection{Epidemic threshold}

For the most general case, in the stationary state ($\dot\mu_i=\dot\rho_i=0$), the following implicit relation $\mu_i^{\star}=\mu_i^{\star}(\rho_i^{\star})$ can be derived from Eq. (1):
\begin{eqnarray}  
    \mu_i^{\star}&= &\beta \left( 1 - \sum_{j=1}^N \rho_j^{\star} \right) - \frac{D_x}{\rho_i^{\star} } \sum_{j = 1}^N  \ell_{i j} \rho_j^{\star}. 
    \label{eq:M1} 
\end{eqnarray}
Moreover, introducing  Eq. (\ref{eq:M1}) in  Eq. (2) we obtain a close implicit relation for $\rho_i^{\star}$:
\begin{eqnarray}  
    \rho_i^{\star}&= & \frac{- D_x\sum_{j = 1}^N  \ell_{i j} \rho_j^{\star}}{\mu_0-\beta \left( 1 - \sum_{j=1}^N \rho_j^{\star} \right)+ \frac{\alpha}{\gamma} \sum_{j=1}^N \left(\rho_j^{\star}e^{-x_{ij}/\Delta}\right)}. 
    \label{eq:M2}
\end{eqnarray}
Close to the epidemic threshold, the disease becomes endemic
displaying a small incidence: $\rho_i^{\star}=\epsilon^{\star}\ll 1\;\forall i$. This allows a linearization of Eq. (\ref{eq:M2}) whose solution is given by:
\begin{eqnarray}
        \frac{\mu_0 - \beta}{D_x} \epsilon_i^{\star} &=& - \sum_{j = 1}^N l_{ij} \epsilon_j^{\star}.
    \label{eq:M3}
\end{eqnarray}
Eq. (\ref{eq:M3}) defines an eigenvalue problem where the epidemic threshold $\beta^c$ (the minimum $\beta$ value so that the above expression holds) is related to the minimum eigenvalue $\Lambda_{\text{max}}$ of the Laplacian $\mathbf{L}$ as:
\begin{eqnarray}
        \beta^c &=& \mu_0 + D\Lambda_{\text{min}}(\mathbf{L}).
        \label{eq:M4}
\end{eqnarray}
Considering that by definition the Laplacian matrices have semipositive eigenvalue spectrum with at least one eigenvalue equal to zero, then $\Lambda_{\text{min}}(\mathbf{L})=0$ and the epidemic threshold of the SIS model is recovered:
\begin{eqnarray}
        \beta^c &=& \mu_0 \Leftrightarrow R_0^c=1.
        \label{eq:M5}
\end{eqnarray}

\newpage

\section{Genotype re-emergence}

The re-emergence of a single genotype is determined by the waning immunity time scale in relation to the other dynamical processes. As demonstrated in Supplementary Fig. \ref{fig:S4}.a, if the waning time scale is comparable to the acquisition one (set here to $\alpha=0.03$), then genotypes coexist since no effective immunity is obtained. However, when the waning process occurs at a time scale that is slower, the prevalence of each genotype decreases after each epidemic wave. Nevertheless, the genotype eventually re-emerges due to the loss of immunity in the population. Following several re-emergences, the epidemic trajectory ultimately converges to a steady, stationary prevalence. 

Furthermore, Supplementary Fig. \ref{fig:S4}.b-e shows the impact of the mutation rate on dynamics, illustrating how, in cases of elevated mutation rates, diffusion becomes the primary driver of antigenic space exploration. Conversely, at lower mutation rates, the acquired immunity towards a genotype leads to the prevalence of adjacent genotypes in the population (antigenic escape). Note that a linear chain of 20 nodes has been used as the genotype network in Supplementary Fig. \ref{fig:S4}.

\begin{figure*}[h]
\centering\includegraphics[width=1\linewidth]{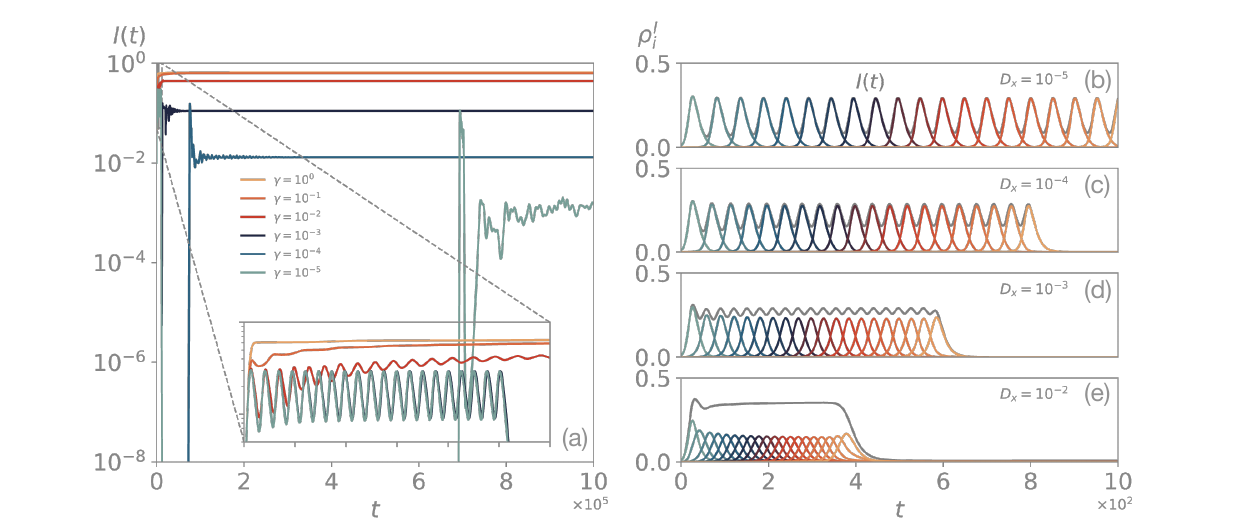}
\caption{\justifying \textbf{Genotype re-emergence and antigenic space exploration} (a) Genotype re-emergence by considering a linear chain of 20 nodes as genotype network. The parameters take the following values: $\beta=0.3$, $\mu_0=0.1$, $\alpha=0.03$, $\Delta=0$ and $D_x=10^{-5}$. (b) Diffusion versus antigenic escape as main driver of viral persistence in a linear chain of 20 genotypes. The  parameters take the following values: $\beta=0.3$, $\mu_0=0.1$, $\alpha=0.03$, $\Delta=0$ and $\gamma=10^{-5}$.}
\label{fig:S4}
\end{figure*}

\newpage

\medskip

\section{Trajectories with transitory chaotic-like behavior}

In case immunity wanes ($\gamma\neq0$), epidemic trajectories can display transitory chaotic-like behaviors, even in simple representations of the antigenic space, such as a linear chain of 20 nodes. Afterwards, they converge into their fixed state ($I^{\star},R^{eff\,\star}$), being $R^{eff\,\star}=n^{-1}\sum_{i=1}^{n}R^{eff\,\star}_i$. This is exemplified in Supplementary Fig. \ref{fig:S5}.a, while in Supplementary Fig. \ref{fig:S5}.b we highlight a section of the trajectory displaying this transitory behavior.

\begin{figure*}[h]
\centering\includegraphics[width=1\linewidth]{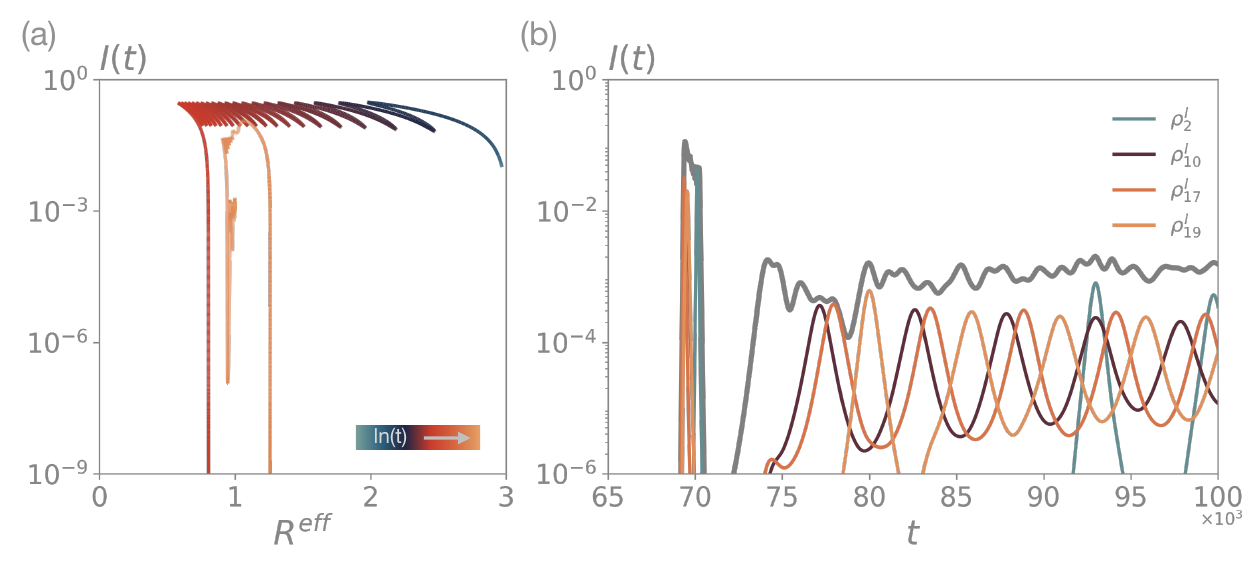}
\caption{\justifying \textbf{Trajectories with transitory chaotic-like behavior} (a) Epidemic trajectory represented in the ($I^{\star},R^{eff\,\star}$)-space. (b) Section of the trajectory with the dynamics of some particular genotypes highlighted. The utilized genotype network is a linear chain of 20 nodes, and the  parameters take the following values: $\beta=0.3$, $\mu_0=0.1$, $\alpha=0.03$, $D_x=10^{-5}$, $\Delta=0$ and $\gamma=10^{-5}$.}
\label{fig:S5}
\end{figure*}

\newpage

\section{Structural properties of genotype networks}

\begin{figure}[H]
\centering\includegraphics[width=0.85\linewidth]{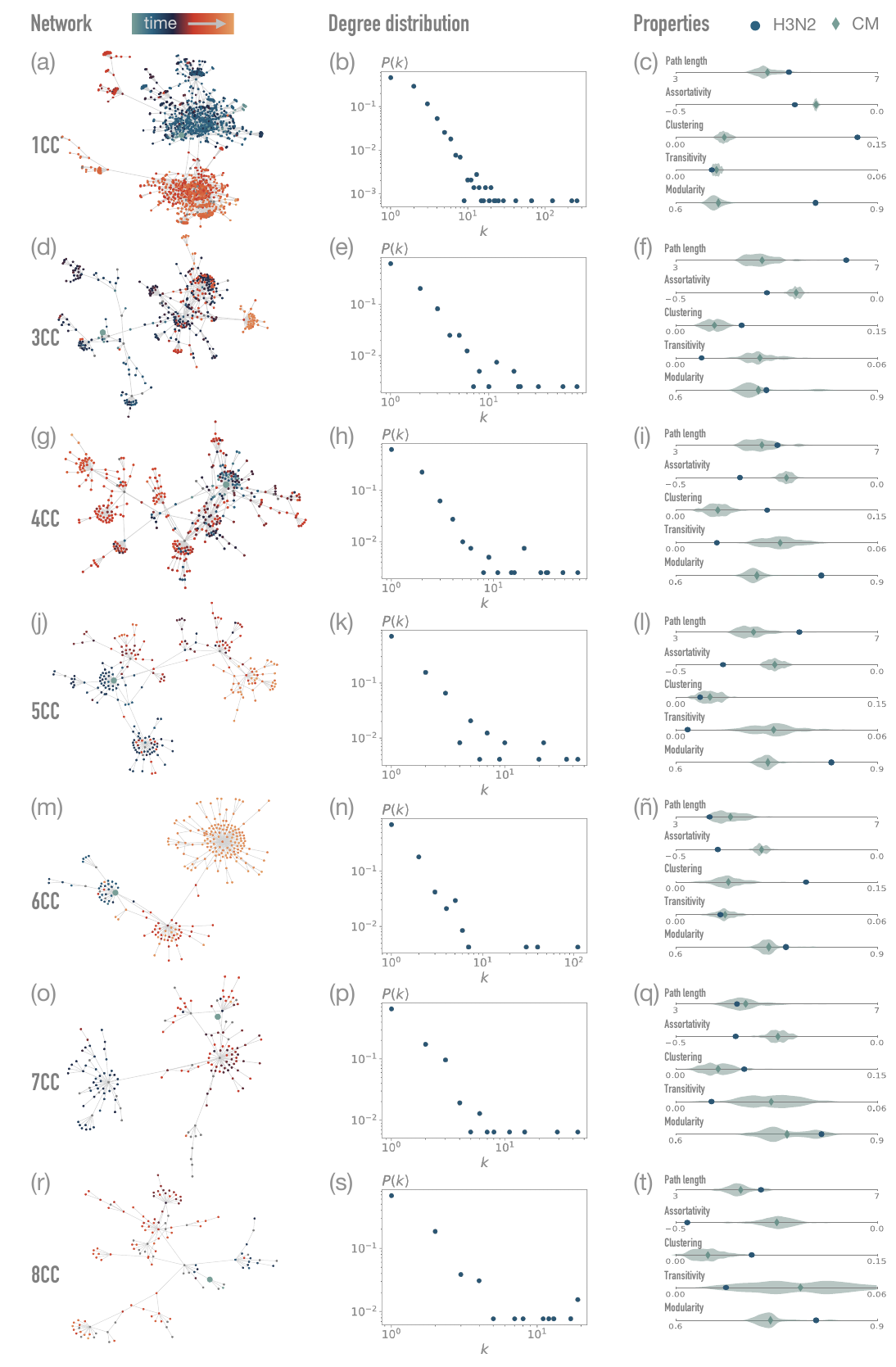}
\caption{\justifying \textbf{Structural properties of genotype networks}. Extension of Fig. 2 of the main text. Full description of the figure in the Supplementary Information text.}
\label{fig:S1}
\end{figure}

Supplementary Fig. \ref{fig:S1} extends the analysis performed in Fig. 2 for the second connected component of the Influenza A genotype network \cite{williams2022immunity} to the remaining first eight largest connected components. Each row corresponds to a different connected component. The left-hand plot for each component presents the network, with the color code denoting the year of sampling. Each plot in the middle column displays the degree distribution of the corresponding component. Finally, the plots in the right column showcase the values of some structural properties of each component. The purple dots correspond to the values of the properties in the real network, while the violin plot corresponds to the values obtained for an ensemble of 100 networks sharing the same degree sequence. Further details on the construction can be found in Methods.

\newpage

\newpage

\section{Dynamical trade-off between cross-immunity and genotype swarms}

\begin{figure*}[h]
\centering\includegraphics[width=1\linewidth]{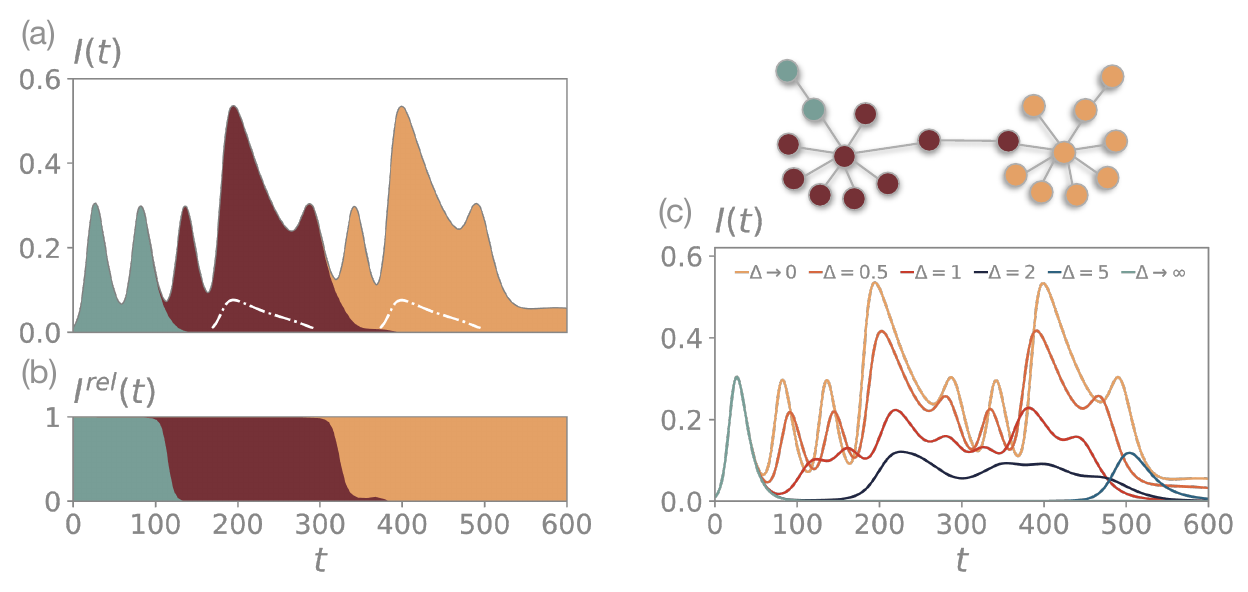}
\caption{\justifying \textbf{Antigenic space shapes epidemic trajectories.} (a) Epidemic trajectory with the adjacent structure as genotype network. Withe curves correspond to the evolution of one of the leaves in each of the swarms. The color corresponds to the regions of the network. $\beta=0.3$, $\mu_0=0.1$, $\alpha=0.03$, $\gamma=0$, $D=10^{-5}$ and $\Delta=0$ (b) Relative prevalence for each of the three regions of the structure. (c) Prevalence evolution in terms of the length of cross immunity $\Delta$. The remaining parameters are those of panels (a)-(b).}
\label{fig:S7}
\end{figure*}

Cross-immunity complements acquired natural immunity to neutralize viral load, and the relationship between the characteristic cross-immunity length and the extension of the antigenic space determines its relevance. Supplementary Fig. \ref{fig:S7}.a illustrates the epidemic trajectory on the adjacent synthetic networks. As shown in Fig. 3.b of the article, the magnitude of the epidemic waves of those genotypes that are identical, from the structure point of view, is similar. If there are swarms of genotypes formed by several leaves around a successful hub, an increase in prevalence is observed. Furthermore, in Supplementary Fig. \ref{fig:S7}.b, the relative prevalence of each of the swarms is shown according to the color-code, to better understand the exploration of the antigenic space.

As shown in Supplementary Fig. \ref{fig:S2}.c, the existence of cross-immunity results in a reduced penetration of genotypes within the population. However, as illustrated in Supplementary Fig. \ref{fig:S7}.c, the presence of mutant swarms can enable the virus to reach comparable prevalences to the initial outbreak. The trade-off between the magnitude of the swarms and the characteristic cross-immunity length thus determines the global prevalence of the epidemic trajectory.

\newpage

\section{Immune response dynamics gives relevance to the differences in structure}

The synthetic genotype networks utilized in Fig. 3 are a lattice, an homogeneous concatenation of star-like genotype swarms and an heterogeneous one. In the homogeneous concatenation of star-like genotype swarms, each swarm has 40 leaves with a node as the nexus between stars. In contrast, in the heterogeneous structure, the number of leaves varies between stars. Finally, the lattice is a homogeneous concatenation of 3-nodes layers. All the structures can be found in the following GitHub repository: https://github.com/santiagolaot/SIMS-model, and their degree distributions are presented in Supplementary Fig. \ref{fig:S6}.a.

Moreover, this section complements Fig. 3 of the main text by showing in Supplementary Fig. \ref{fig:S6}.b that, in absence of immune response dynamics (dashed lines, $\alpha=0$) all three structures present analogous macroscopic dynamics.

\begin{figure*}[h]
\centering\includegraphics[width=0.98\linewidth]{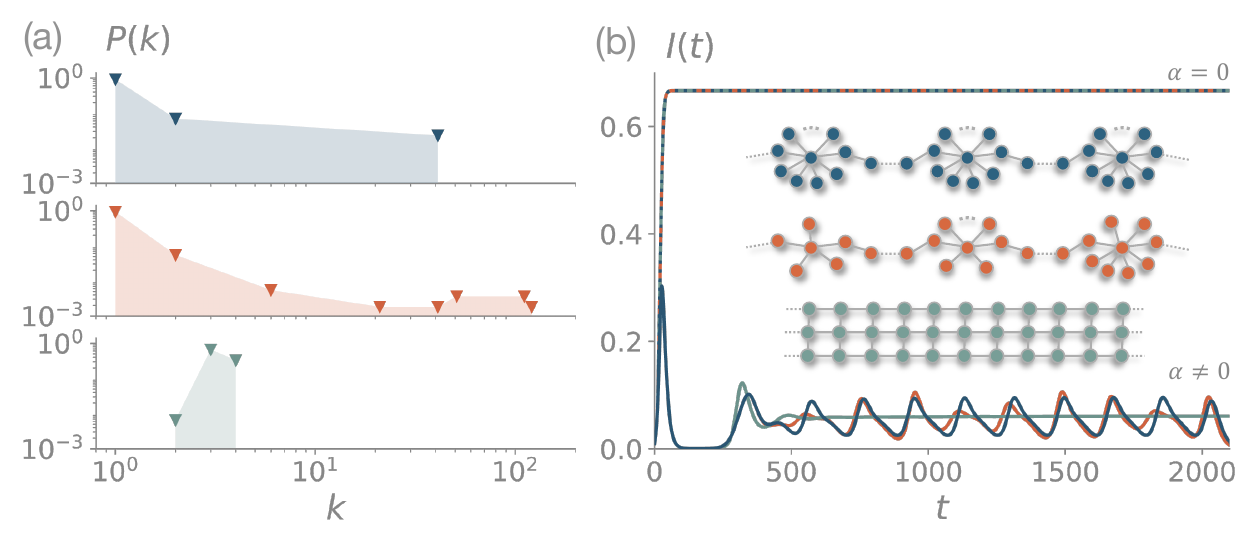}
\caption{\justifying\textbf{Synthetic genotype networks.} (a) Degree distributions of the three synthetic structures utilized, namely a homogeneous concatenation of stars (top), a heterogenoeus one (middle) and the homogeneous lattice (bottom). (b) The purple lines show the epidemic trajectories in a homogeneous concatenation of star-like genotype networks, the blue lines show the epidemic trajectories in a lattice, and the yellow lines show the epidemic trajectories in a heterogeneous concatenation of star-like genotype networks. While continuous lines reproduce Fig. 3.b of the article ($\alpha=0.03$), dashed lines present the $\alpha=0$ scenario. In both cases $\beta=0.3$, $\mu_0=0.1$, $\gamma=0$, $D_x=10^{-5}$ and $\Delta=3$.}
\label{fig:S6}
\end{figure*}

\newpage

\section{Genotype swarm alternation and re-emergence in real-world genotype networks}

As discussed in Supplementary Fig. \ref{fig:S6}, the length of the cross-immunity shapes epidemic trajectories. In Supplementary Fig. \ref{fig:S8}.a (the display of which replicates Fig. 4 in the main text), we show for the second largest connected component of the Influenza A genotype network that $\Delta\rightarrow0$ leads to a genotype swarm alternation towards the convergence to the stationary state. Due to the large number of nodes and the complexity of the genotype structure, individual genotypes do not necessarily reach $R_i^{eff}<1$ after the first wave, and therefore they dominantly prevail several times until the absorbent stationary state is reached (note that $\gamma=0$). In this regime, the observed alternation is due to competition between different genotype swarms. While $\langle R_{i\in{\textcolor{black}{\blacksquare}}}^{eff}\rangle>1$, where $\textcolor{black}{\blacksquare}\in\left\{\textcolor{color1}{\blacksquare},\textcolor{color6}{\blacksquare},\textcolor{color5}{\blacksquare},\textcolor{color4}{\blacksquare},\textcolor{color2}{\blacksquare},\textcolor{color3}{\blacksquare}\right\}$, partial immunity is still acquired due to $\alpha$, and the pool of susceptibles for the dominant genotype swarm narrows. For example, after the first impact of the teal genotype swarm, $\langle R_{i\in{\textcolor{color1}{\blacksquare}}}^{eff}\rangle$ becomes smaller than $\langle R_{i\in{\textcolor{black}{\blacksquare}}}^{eff}\rangle$, where $\textcolor{black}{\blacksquare}\in\left\{\textcolor{color6}{\blacksquare},\textcolor{color5}{\blacksquare},\textcolor{color4}{\blacksquare},\textcolor{color2}{\blacksquare},\textcolor{color3}{\blacksquare}\right\}$, which causes the other swarms to overtake its prevalence in the population.

In Supplementary Fig. \ref{fig:S8}.b we show, for the second connected component of the Influenza A network, the re-emergence of the genotype swarms as a consequence of waning immunity ($\gamma=10^{-4}$, and keeping $\Delta=3$ as in Fig. 4 of the main text). As expected, trajectories with transitory chaotic-like behavior are obtained, as in Supplementary Fig. \ref{fig:S4}.

\begin{figure*}[h]
\centering\includegraphics[width=0.98\linewidth]{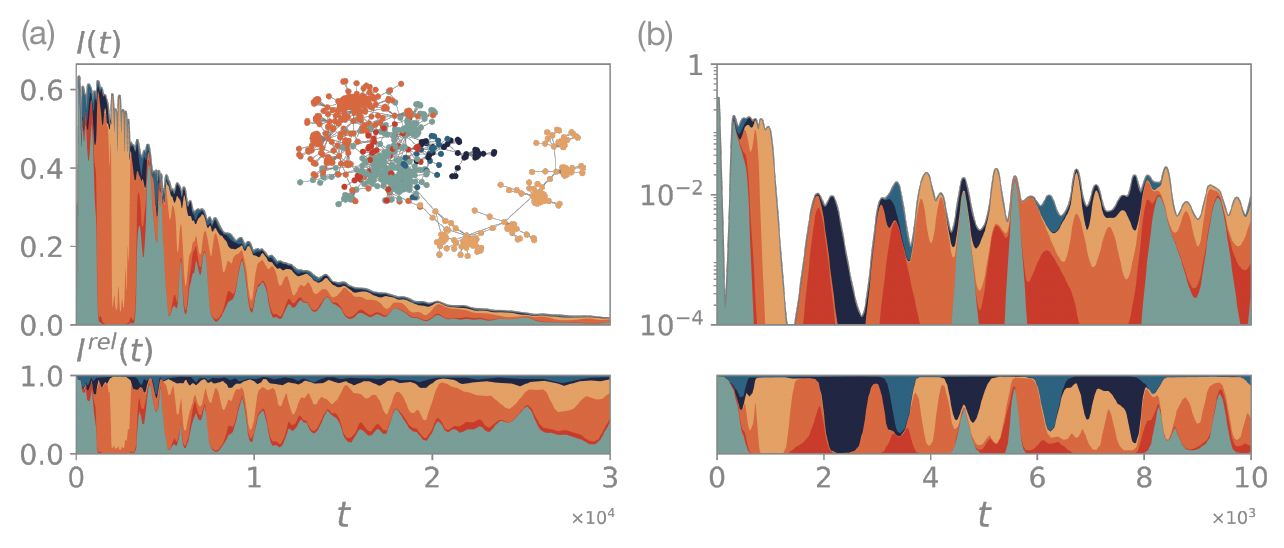}
\caption{\justifying\textbf{Alternation and re-emergence of genotype swarms in real-world genotype networks.} Absolute and relative epidemic trajectories for the second largest connected component of the INFV A (H3N2) network \cite{williams2022immunity}. Panel (a) shows the alternation of genotype swarms for $\Delta\rightarrow0$ and $\gamma=0$, and panel (b) shows the genotype swarm re-emergence for $\Delta=3$ and $\gamma=10^{-4}$. In all panels, $\beta=0.3$, $\mu_0=0.1$, $\alpha=0.03$ and $D_x=10^{-5}$. As an inset, in panel (a) we represent the network with the colors corresponding to the mutant swarms of nodes on each of the structures.}
\label{fig:S8}
\end{figure*}

\newpage

\clearpage


\end{document}